\documentclass[twocolumn]{aastex63}
\usepackage{amsmath}
\usepackage{amssymb}
\usepackage{bm}
\usepackage{graphicx}
\usepackage{color}
\usepackage{float}

\definecolor{darkgreen}{rgb}{0,0.35,0}
\newcommand\as{\bgroup\markoverwith{\textcolor[rgb]{.5, 0, .6}{\rule[0.5ex]{8pt}{1.5pt}}}\ULon}

\shorttitle{BBH in AGN Disks}
\shortauthors{Li et al.}

\begin{document}

\title{Orbital evolution of binary black holes in active galactic nucleus disks: a disk channel for binary black hole mergers?}

\author[0000-0002-7329-9344]{Ya-Ping Li}
\affiliation{Theoretical Division, Los Alamos National Laboratory, Los Alamos, NM 87545, USA}
\author[0000-0001-8291-2625]{Adam M. Dempsey}
\affiliation{Theoretical Division, Los Alamos National Laboratory, Los Alamos, NM 87545, USA}
\author[0000-0002-4142-3080]{Shengtai Li}
\affiliation{Theoretical Division, Los Alamos National Laboratory, Los Alamos, NM 87545, USA}
\author[0000-0003-3556-6568]{Hui Li}
\affiliation{Theoretical Division, Los Alamos National Laboratory, Los Alamos, NM 87545, USA}
\author[0000-0001-5550-7421]{Jiaru Li}
\affiliation{Theoretical Division, Los Alamos National Laboratory, Los Alamos, NM 87545, USA}

\correspondingauthor{Ya-Ping Li}
\email{leeyp2009@gmail.com}
\date{\today, accepted by ApJ}

\begin{abstract}
We perform a series of high-resolution 2D hydrodynamical simulations of equal-mass binary black holes (BBHs) embedded in active galactic nucleus (AGN) accretion  disks to study whether these binaries can be driven to merger by the surrounding gas.  
We find that the gravitational softening adopted for the BBH has a profound impact on this result.
When the softening is less than ten percent of the binary separation, we show that, in agreement with recent simulations of isolated equal-mass binaries, prograde BBHs expand in time rather than contract.
Eventually, however, the binary separation becomes large enough that the tidal force of the central AGN disrupts them.
Only when the softening is relatively large do we find that prograde BBHs harden.
We determine through detailed analysis of the binary torque, that this dichotomy is due to a loss of spiral structure in the circum-single disks orbiting each black hole when the softening is a significant fraction of the binary separation. 
Properly resolving these spirals -- both with high resolution and small softening -- results in a significant source of binary angular momentum. 
Only for retrograde BBHs do we find consistent hardening, regardless of softening, as these BBHs lack the important spiral structure in their circum-single disks. 
This suggests that the gas-driven inspiral of retrograde binaries can produce a population of compact BBHs in the gravitational-wave-emitting regime in AGN disks, which may contribute a large fraction to the observed BBH mergers. 
\end{abstract}

\section{Introduction}

Since the first discovery of a transient gravitational-wave signal by the two detectors of the Laser Interferometer Gravitational-Wave Observatory (LIGO;\citealt{Abbott2016}), more than fifty mergers have been discovered during LIGO/Virgo’s first three observing runs \citep{Abbott2019PhRvX,Abbott2020_GWTC2}.  Active galactic nucleus (AGN) disks have been proposed as promising locations for producing some of the detected stellar mass binary black hole (BBH) mergers \citep{McKernan2012,McKernan2014,McKernan2018,McKernan2019,Bartos2017,Stone2017,Leigh2018,Secunda2019,Secunda2020a,Yang2019a,Yang2019b,Grobner2020,Ishibashi2020,Tagawa2020,Tagawa2020b,LigoVirgo2020}.
Other favored scenarios include isolated binary evolution in a binary star system \citep[e.g.,][]{Belczynski2010,Mandel2016,Belczynski2016}, dynamical evolution of triple or quadruple systems \citep[e.g.,][]{Antonini2017,Liu2017,LiuLai2019,Michaely2019,Fragione2019}, and dynamical formation in which the black holes (BHs) undergo a chance encounter in a dense stellar environment such as galactic nuclei, globular clusters or open clusters \citep[e.g.,][]{OLeary2009,Wang2016,Banerjee2017,Rodriguez2018,Fernandez2019}.

An AGN disk is a favorable
location for BBH mergers
because the surrounding gaseous disks may harden existing BBHs
\citep{McKernan2014,Bartos2017,Stone2017,Yang2019a}. 
AGN disk-assisted BH mergers have distinct properties that
could differentiate them from other merger channels.
These include heavier BBH mergers \citep{Yang2019a,LigoVirgo2020}, a spatial correlation with AGN host galaxies \citep{Bartos2017a}, large spin magnitudes \citep{McKernan2012}, and possible electromagnetic counterparts due to the BHs accreting from the surrounding gaseous disk \citep{Stone2017,Bartos2017,McKernan2019,Graham2020}.

The formation of BBHs within the AGN disk can be due to 
migration traps \citep{McKernan2012}. As BHs accumulate
near these traps, they can form BBH systems \citep{Bellovary2016,Secunda2019}. In addition, BBHs can be formed via isolated evolution of the stellar binary systems \citep{Mandel2016},  or formed dynamically through the chance encounters of BHs in galactic nuclei \citep{OLeary2009}, or formed within the self-gravitating disk itself \citep{Stone2017}.

Most works on the AGN-assisted channel adopt semi-analytic results for the drag force in a uniform gaseous background medium \citep{Kim2007,Kim2008} or from a circum-binary disk (CBD) \citep{Goldreich1980,Cuadra2009,Roedig2012,DOrazio2013} to conclude that gas surrounding the BBH facilitates its merger. This requires confirmation from high resolution numerical simulations since the real situation for an embedded binary in an AGN disk is much more complicated.

The expectation of BBH coalescence stems from early hydrodynamical simulations of isolated binaries \citep[e.g.,][]{MacFadyen2008,Shi2012}.
However, these simulations did not include the binary itself in the computational domain and so missed the important source of angular momentum coming from the circum-single disks (CSDs).
Later work showed that including and appropriately resolving CSDs in the simulation causes isolated binaries to {\it expand} rather than {\it contract} \citep{Tang2017,Munoz2019,Munoz2020,Duffell2020}. Similar results have been obtained from 3D hydrodynamical simulations \citep{Moody2019}.
Moreover, early simulations of isolated binaries were typically evolved for a short amount of time. 
This was shown to cause transient hardening whereas in steady-state the binary expands \citep{Miranda2017}.

Recently, there have been some isolated binary simulations which showed that the binary can still be hardened by the gaseous disk with a small binary mass ratio \citep{Duffell2020,Derdzinski2020} or a cold disk (i.e., small disk scale height, \citealt{Tiede2020,Heath2020}). \citet{Chen2020} found analytically that the radiation field from AGN also causes the orbits of the accreting super-massive black hole binaries (SMBHBs) with a small mass ratio to shrink by the ``Poynting-Robertson" drag effect.

To our knowledge, there has only been one work that has used 2D hydrodynamical simulations to study an embedded binary star in an AGN disk \citep{Baruteau2011}.
While the outcome of their simulations agreed with the semi-analytic and early isolated binary results, the adopted resolution and gravitational softening could not adequately resolve the innermost regions (i.e., CSD regions) around each star.
Therefore, the AGN disk channel for BBH and SMBHB mergers remains an active research topic.   

In this work, we have performed a series of high-resolution 2D hydrodynamical simulations of BBHs embedded in thin accretion disks. 
The resolution and gravitational softening in these simulations are chosen so that the CSDs around each BH are properly resolved.
Our primary goal is to examine whether or under what conditions a BBH will harden in an AGN disk. 
Motivated by the isolated binary simulations, we quantitatively explore the effects of the CSD structures on the binary evolution.

The paper is organized as follows. In Section~\ref{sec:method} we introduce the numerical setup of our simulations, and in Section~\ref{sec:results} 
we present our numerical results. We explore the binary evolution for different gravitational softening scales and prograde/retrograde orbits.  Finally, we summarize the main finding of our paper, and then discuss the limitations of our present work and the future direction in Section~\ref{sec:conclusions}. 

\section{Method}\label{sec:method}

%\subsection{Simulation Setup}
We numerically solve the 2D continuity and Navier-Stokes equations using the hydrodynamical code \texttt{LA-COMPASS} \citep{Li2005,Li2009}, 
to study a BBH embedded in a gaseous disk that surrounds a super-massive black hole  (SMBH). 
For simplicity, only one BBH is included in the disk. 

In our simulations, we use a geometrically thin and non-self-gravitating disk. We choose a 2D cylindrical coordinate system $(r,\varphi)$, with the origin located at the position of the central SMBH.

\subsection{Initial Conditions}

We adopt a simple model in which the disk temperature follows $T_{\rm disk}\propto r^{-1}$.
With this,  the disk aspect ratio $h_{0}\ (\equiv H/r)$ is constant throughout the whole disk and we choose $h_{0}=0.08$.
Here $H$ is the disk scale height. The disk surface density has an initial profile of 
\begin{equation}
    \Sigma(r)=\Sigma_{0}\left(\frac{r}{r_{0}}\right)^{-0.5},
    \label{eq:sigmar}
\end{equation}
where $\Sigma_{0}$ is the gas surface density at $r_{0}$, and $r_{\rm 0}$ is the initial radius of the binary circular orbit for the BBH center of mass (COM).
The 2D velocity vector of the gas is $\bm{v}=\left(v_{r}, {v}_{\varphi}\right)$, and angular velocity is $\Omega=v_{\varphi}/r$.
We assume an $\alpha$-prescription for the gas kinematic viscosity $\nu=\alpha H^2 \Omega$ \citep{Shakura1973}.
Such a combination of $\Sigma$, $T_{\rm disk}$, and $\nu$ ensures an initial steady state of disk accretion.

We choose the unit of length to be $r_{0}$, and the unit of mass to be the SMBH mass, $M_{\rm smbh}$.
We set $\Sigma_{0} = 10^{-4}\ M_{\rm smbh}/r_0^2$, which corresponds to a total disk mass of $10^{-3}\ M_{\rm smbh}$.
With these values, the gas self-gravity is negligible.

\subsection{Gravitational Interaction}

We initialize an equal mass BBH on a circular orbit with a semi-major axis of $a_{0} = \kappa R_{\rm H}$, where $R_{\rm H} = (2 q/3)^{1/3} r_0$ is the BBH Hill radius, and $q$ is the mass ratio of one BH to the SMBH.
We keep $\kappa < 1$, so that the SMBH does not strongly affect the gas dynamics in the vicinity of the BBH.
The binary orbital period is $P_{\rm bin} = \sqrt{\kappa^3/3}\ P_{\rm com}$ in units of the binary's COM orbital period around the SMBH, $P_{\rm com}$.

To the gas momentum equation, we add the gravitational potential,
\begin{eqnarray}
    \phi = &-&\frac{G M_{\rm smbh}}{\left|\bm{r}\right|} \nonumber \\
    &+& \sum_{i=1}^{2} \left[-\frac{G qM_{\rm smbh}}{(\left|\bm{r}_{{\rm bh},i}-\bm{r}\right|^2+\epsilon^2)^{1/2}}
    + q \Omega_{{\rm bh},i}^{2} \bm{r}_{{\rm bh},i} \cdot \bm{r}
    \right].
    \label{eq:potential}
\end{eqnarray}
The location and orbital frequency of each BH with respect to the SMBH are given by $\bm{r}_{{\rm bh},i}$ and $\Omega_{{\rm bh},i}$, where $i=1,2$ labels each component of the BBH. $G$ is the gravitational constant.
The first term is the potential due to the central SMBH, while the first term in the bracket is the direct potential from each BH.
The second term in the bracket is the indirect potential arising from our choice of coordinate system.
We apply a gravitational softening, with length scale
$\epsilon$, to each BH potential. 
Based on the isolated binary simulations, a small gravitational softening scale (i.e., $\ll a_{0}$) is needed to resolve the CSD in the vicinity of each BH. 
Here, we mainly focus on a softening of $0.08\ a_{0}$, which is comparable to the softening scale in isolated binary simulations \citep[e.g.,][]{Munoz2019,Munoz2020,Tiede2020}. To better understand the influence of the softening on our results, we also explore a range of values.

Applying such a small softening requires a super resolution around each BH to resolve their CSDs.
For typical stellar mass BH and SMBH masses, $q\simeq10^{-8}\sim10^{-5}$ and $R_{\rm H} \simeq 0.0015\sim0.032\ r_{0}$.
Given that the size of the CSD region is much smaller than $R_{\rm H}$, resolving this region for realistic parameters is unfeasible. 
Because of this, we adopt a larger mass ratio of  $q=2\times10^{-3}$, disk viscosity $\alpha=6.35\times10^{-3}$, and disk scale height $h_{0}=0.08$. 
This setup could maintain the same gap profile with the more realistic system of $q=1.4\times10^{-5}$, $\alpha=0.01$ and $h_{0}=0.01$ \citep{Lin1993,Crida2006,Baruteau2011}.

\subsection{Gas Accretion} 

Our BH accretion prescription gradually removes mass and momentum from the gas within a distance of $r_{\rm acc} = 0.0016\ r_0$ to each BH, which is comparable to the BH softening scale.
Material that is accreted by a BH is not added to its mass or momentum.
We fix the time scale of this removal to  $\tau_{{\rm acc},i}^{-1} = 5\ \Omega_{{\rm bh},i}$ unless otherwise stated.
We have tested that this choice of $\tau_{{\rm acc},i}$ -- even $\tau_{{\rm acc},i}\rightarrow \infty$ -- does not change our main results significantly (see also \citealt{Li2021} for the effect of $\tau_{\rm acc}$ on the accretion rate). 
We discuss other removal rates in Section \ref{sec:accretion}.

\subsection{Resolution and Boundary Conditions}

We resolve the disk with a uniform radial grid of $n_{r}=2048$ points in a radial domain between $[0.4,2.0] r_0$, and a uniform azimuthal grid with $n_{\phi}=8192$ points. 
With such a high resolution, the BBH Hill radius can be resolved with $\sim140$ cells.
We also test that a larger radial domain with $r\sim$ $[0.4,4.0]r_0$ in a logarithmic radial grid of $n_{r}=3072$ does not have a significant impact on the binary dynamics. 

At both boundaries we fix the gas to its initial axisymmetric configuration.
To facilitate this, we gradually remove any waves near both boundaries with wave killing zones \citep{deValborro2006}. 

Our simulations are performed in two steps. The BBH is
first held on a fixed circular orbit for 500 orbits, so that the disk can reach a quasi-steady state.
During this time, the BBH opens a gap of similar depth and shape to an equivalent mass single planet. After 500 orbits, the force from the disk on the BBH is included and the binary can dynamically evolve. 
For the entirety of the simulation, we do not allow the time step to exceed $1/400$ of the binary's current internal orbital period.
This ensures that the dynamics in the CSDs is adequately resolved in time \citep{Cresswell2006,Baruteau2011}.

\section{Numerical results} \label{sec:results}

\begin{table}[t]
  \begin{center}
  \caption{\bf Model parameters and binary evolution }\label{tab:para}
  \begin{tabular}{lcccc|c}
     \hline\hline
     % after \\: \hline or \cline{col1-col2} \cline{col3-col4} ...
     Model & $a_{0}$ & ${\epsilon}$ &  accretion   & prograde/ &   Comments\\
                & $(r_{0})$ & $(a_{0})$ &    &  retrograde\\
     \hline
     
     A & 0.02 & 0.08 & on & prograde & outspiraling\\
     Ar & 0.02 & 0.08 & on$^a$ & prograde & outspiraling\\
     B & 0.02 & 0.4 & on & prograde & inspiraling\\
     C & 0.02 & 0.2 & on & prograde & outspiraling\\
     D & 0.02 & 0.08 & off & prograde & outspiraling\\
     %D & 0.02 & 0.08 & acc2 & prograde & outspiraling\\
     E & 0.04 & 0.04$^{b}$ & on & prograde & outspiraling\\
     F & 0.04 & 0.2$^{b}$ & on & prograde & inspiraling\\
     G & 0.04 & 0.04$^{b}$ & on & retrograde & inspiraling\\
     H & 0.04 & 0.2$^{b}$ & on & retrograde & inspiraling\\
     %\hline
    
     \hline\hline
   \end{tabular}
   \end{center}
   \tablecomments{
   $^{a}$ A smaller removal rate of $0.1\Omega_{{\rm bh},i}$ and a larger radial domain of $r\sim[0.3,4]$  for the BBH is adopted to test the effect of binary accretion. \\
   $^{b}$ We keep the same physical length of gravitational softening as Model A or B.
   For the first five Models (A, Ar, B, C \& D), $a_{0} = 0.18\ R_{\rm H}$, while for the remaining Models $a_{0}=0.36\ R_{\rm H}$.
   }
   
\end{table}

\begin{figure*}[htbp]
\centering
\includegraphics[width=0.9\textwidth,clip=true]{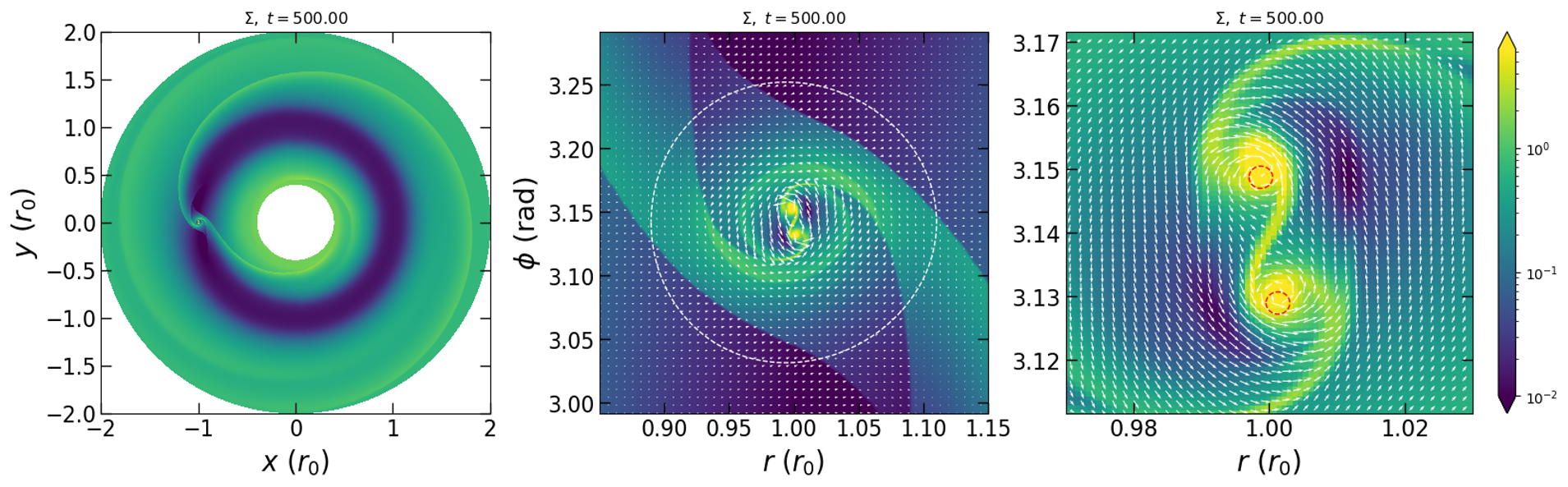}
\caption{Snapshots for the gas surface density in Model A at $500$ COM orbits. From left to right, each panel zooms in closer to the BBH.
The dashed white circle shows the BBH Hill radius.
Overlaid arrows show the gas velocity streamline in the co-moving frame of the BBH COM. The dashed red circles in the right panel correspond to the softening scale for each BH.}
 \label{fig:sigma_snapshot}
\end{figure*}

Table~\ref{tab:para} provides an overview of all of our simulations.
We first study the cases where the binary and disk angular momenta are aligned.  This is the prograde case shown in Table~\ref{tab:para}.

\subsection{Gas Structures}

As a typical case, we choose Model A to show the gas surface density in Figure\ \ref{fig:sigma_snapshot}. 
On scales larger than $R_{\rm H}$, the disk is nearly indistinguishable from a single large mass planet \citep[cf.][]{Dempsey2020a}.
On smaller scales, the gas streamlines reveal that gas circulates around the BBH COM. Based on this, there is a clear ``CBD" region just inside $R_{\rm H}$. This is different from the viscously controlled CBD in isolated binary simulations \citep[e.g.,][]{Tang2017,Munoz2019}, as
the existence of the central SMBH drives large-scale spiral arms that feed the CBD.

Zooming even closer in to the binary shows clear CSDs. With such a small softening for Model A, the CSD around each BH is properly resolved by our high resolution, and contains several spiral arms. 
In particular, the inter-spiral arm connecting the two BHs is quite prominent, as are the two trailing spiral arms that connect to the large scale spirals through the CBD region. 
There is not a sharp cavity around the CSD region as shown in many isolated binary simulations \citep[e.g.,][]{Tang2017,Munoz2019}, which is probably due to the shallow depth of the gap carved by the BBH.

\begin{figure}[htbp]
\centering
\includegraphics[width=0.45\textwidth,clip=true]{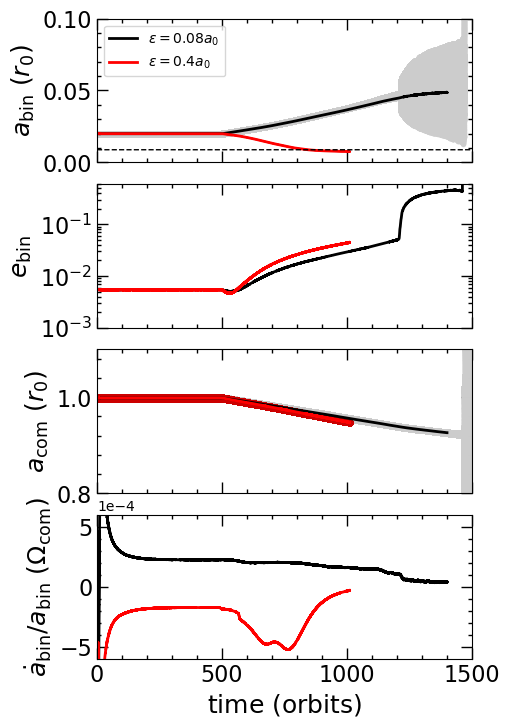}
\caption{Binary evolution for Models A (black), and B (red). The first panel shows the time evolution of the BBH instantaneous separation (shaded region) and semi-major axis (solid lines), while the second panel shows the BBH eccentricity.
The third panel plots the time evolution of the BBH COM semi-major axis,
and the last panel shows the rate of change of the BBH semi-major axis due to the disk forces.
The dashed line in the first panel corresponds to the softening scale in Model B. 
In Models A, the BBH expands while increasing its eccentricity, while in Model B the BBH contracts to the softening scale. 
}
 \label{fig:rbin}
\end{figure}

\subsection{Binary Dynamics}

In Figure~\ref{fig:rbin}, we show the time evolution of the BH-BH and BBH-SMBH orbits for Model A.
The orbit of the BH-BH binary is characterized by the semi-major axis $a_{\rm bin}$ and eccentricity $e_{\rm bin}$, while the BBH-SMBH binary has semi-major axis $a_{\rm com}$ and is circular throughout its evolution.
After release, $a_{\rm bin}$ increases with time as shown in the upper panel. 
This is consistent with many recent isolated binary simulations \citep{Miranda2017,Moody2019,Munoz2019,Munoz2020,Duffell2020,Tiede2020}. 
The shaded region shows the instantaneous separation between the two BHs. We can see that after $\sim1250$ COM orbits, the BBH's eccentricity is strongly excited to $e_{\rm bin} \sim 0.5$ until the binary dissolves at $\sim1500$ orbits.

In contrast to Model A, in the large softening case of $0.4\ a_{0}$ (Model B), the binary gradually inspirals at near zero eccentricity until it reaches the softening scale and stalls \citep[as seen in e.g.,][]{Baruteau2011}. 
For this case, the COM migration rate is similar to a single $q=4\times 10^{-3}$ planet \citep[e.g.,][]{Chen2020mig}. 
This suggests that the global migration is mainly controlled by the torque far beyond the binary's CBD  region.

To explore the effect of the binary release on the separation evolution, we show the long-term evolution rate of $\dot{a}_{\rm bin}/a_{\rm bin}$ contributed by the gaseous disk in the fourth panel of Figure~\ref{fig:rbin}, which is defined as
\begin{equation}
    \frac{\dot{a}_{\rm bin}}{a_{\rm bin}}=-\frac{\dot{\varepsilon}_{\rm bin}}{\varepsilon_{\rm bin}},
\label{eq:dotab}
\end{equation}
where  $\varepsilon_{\rm bin}=-GqM_{\rm smbh}/a_{\rm bin}$ is the specific energy of the binary, and 
\begin{equation}
    \frac{d\varepsilon_{\rm bin}}{dt}=(\dot{\bm{r}}_{{\rm bin},1}-\dot{\bm{r}}_{{\rm bh},2})\cdot(\bm{f}_{\rm grav,1}-\bm{f}_{\rm grav,2}) .
\label{eq:dotepb}
\end{equation}
Here, $\bm{f}_{\rm grav,1}$ and $\bm{f}_{\rm grav,2}$ are the gravitational force for each BH from the disk. We have ignored the contribution from accretion since we do not add the accreted mass and momentum to the binary. 
By the time the binary is released, $\dot{a}_{\rm bin}$ is converged in time and there is no impulsive kick upon release.

\begin{figure}[htbp]
\centering
\includegraphics[width=0.45\textwidth,clip=true]{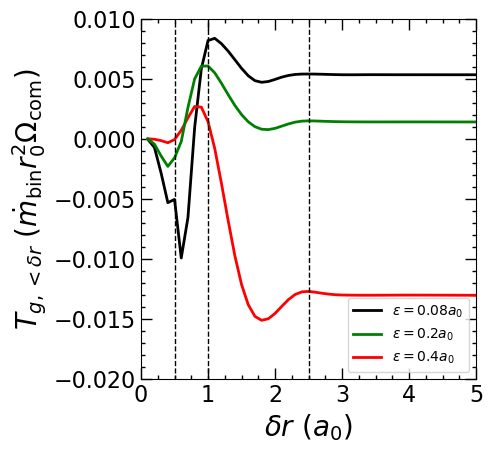}
\caption{Cumulative gravitational torque for the binary evolution as a function of the distance to the binary's COM $\delta r$ for Models A, B, and C. 
The dashed lines correspond to the three torquing regions discussed in the text: the inter-spiral arms, CSDs, and CBD.
Gas outside of $\delta r\sim 2.5\ a_{0}$, provides little to no torque on the BBH. 
The sign of $T_{\rm g}$ for $\delta r \gg 1$ determines whether the BBH hardens ($T_{\rm g} < 0$) or expands ($T_{\rm g} > 0$).
}
\label{fig:torq_r}
\end{figure}

\subsection{Gravitational Torque}
\label{sec:torqden}

\begin{figure*}[htbp]
\centering
\includegraphics[width=0.98\textwidth,clip=true]{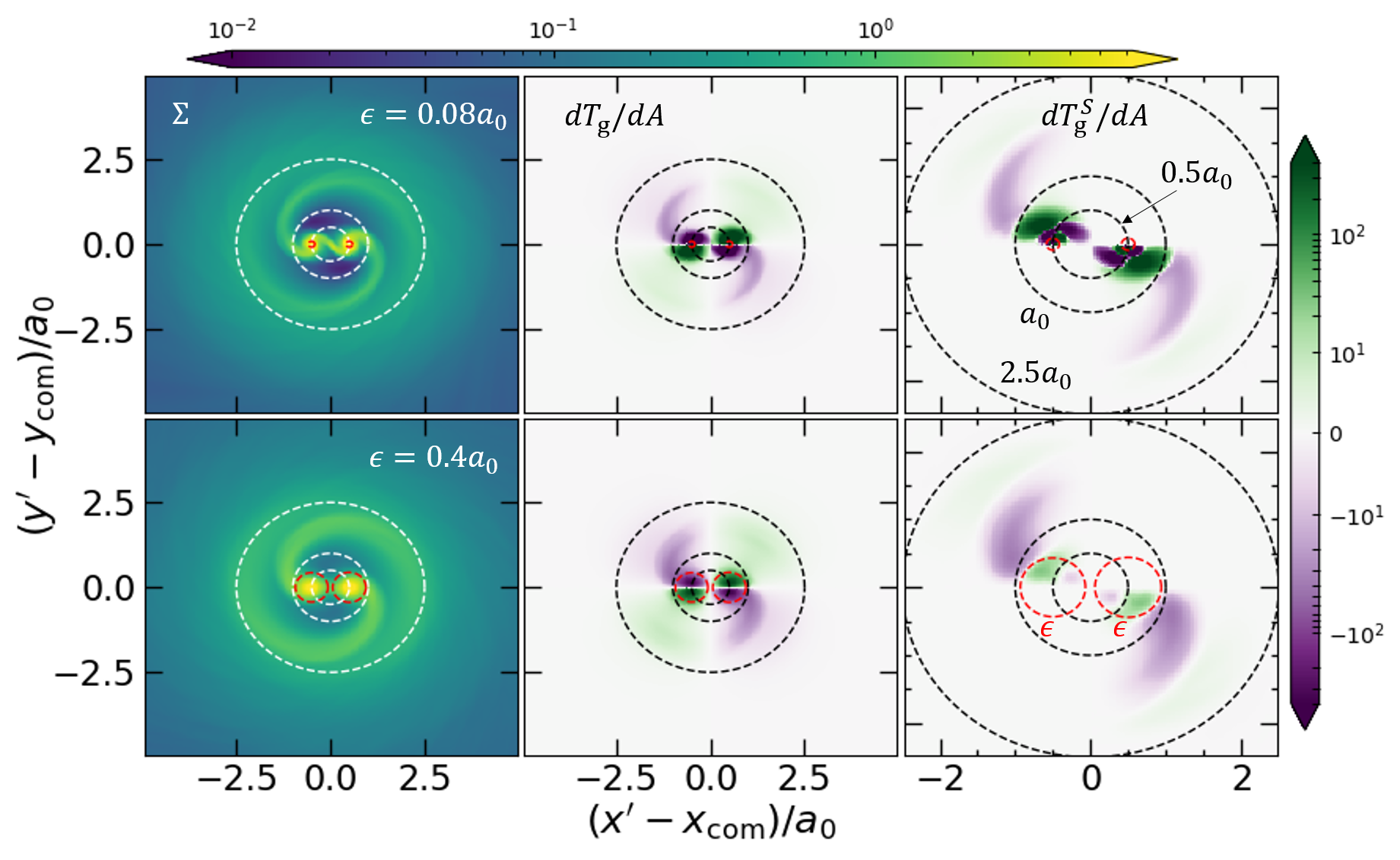}
\caption{Stacked gas surface density (left panels) and specific gravitational torque density $dT_{\rm g}/dA$ (middle and right panels) at $t=500$ orbits for binary's COM frame averaged over 100 snapshots. The coordinate $(x^{\prime}-x_{\rm com},y^{\prime}-y_{\rm com})$ is the frame co-rotating with the BBH with its COM located at $(x_{\rm com},y_{\rm com})$. The summation of $dT_{\rm g}/dA$ (middle panels) over the vertical coordinate gives the symmetric component of the torque density $dT_{\rm g}^{S}/dA$ (right panels), showing in the second and fourth quadrants.  Upper panels: softening scale is $0.08\ a_{0}$ for each BH. Lower panels: softening scale is $0.4\ a_{0}$. The white and black dashed circles at $\delta r=0.5\ a_{0}$, $\delta r=a_{0}$, and $\delta r=2.5\ a_{0}$, in these plots delineate the different zones for the torque spatial decomposition. The red dashed circles denote the corresponding gravitational softening scales.
}
 \label{fig:torqden}
\end{figure*}

In the following, we analyze the torque acting on the binary to understand what causes the binary inspiraling/outspiraling for the different softening lengths. 
To decompose the torque density spatially around the binary, we follow the isolated binary literature and compute the gravitational torque density in three different regions: inter-spiral arms ($\delta r\le 0.5a_{0}$), CSDs ($0.5a_{0}<\delta r\le a_{0}$), and CBD ($\delta r> a_{0}$). 
%outer-spiral arms ($a_{0}<\delta r\le 2.5a_{0}$), 
The torque density from each computational cell acting on the binary is 
$d\bm{T}_{\rm g}/dA=(\bm{r}_{{\rm bh},1}-\bm{r}_{{\rm bh},2})\bm{\times}(\bm{f}_{{\rm grav},1}-\bm{f}_{{\rm grav},2})$.

In Figure \ref{fig:torq_r}, we show the the cumulative radial torque with a time averaging around 500 COM orbits, $T_{\rm g}(<\delta r)= \int_0^{\delta r} (dT_g/dA) dA$ with $\delta r=|r-r_{\rm com}|$, for Models A,B, and C.
All models, regardless of softening, find that gas inside of $0.5\ a_{0}$ and outside of $a_{0}$ torque the BBH down, while gas between $0.5\ a_{0}$ and $a_{0}$ torque the binary up.
However, the amount that each region contributes to the total torque is highly softening dependent. 
As the softening becomes smaller, the total amount of positive torque coming from inside of $a_{0}$ increases until it overcomes the amount of negative torque coming from the CBD region. 
The transition from CBD to CSD dominated angular momentum transfer, and consequently from inspiraling to outspiraling, occurs around a softening of $\epsilon \approx 0.2\ a_{0}$.

To better understand the spatial contribution of the gravitational torque, we follow \citet{Munoz2019} (see also \citealt{Tang2017,Tiede2020}) to reconstruct the 2D torque density distribution around 500 COM orbits. 
We stack the gas surface density $\Sigma$ and torque density $dT_{\rm g}/dA$ in the binary internal co-rotating frame over one binary's COM period with $100$ snapshots, and show the results in Figure~\ref{fig:torqden}. 
The upper and lower panels correspond to the small (Model A) and large (Model B) softening cases, respectively. 
The symmetric component of the torque density with respect to the vertical coordinate, are shown in the right panels. 
The total torque on the BBH can be obtained by summing all four quadrants in the middle panels, or by summing the upper left and lower right quadrants in the right panels.

From the stacked surface density shown in the left panels, 
the {\it coherent} inter-spiral arm and outer spiral arms connecting the two BHs are  quite prominent. 
In contrast, the spiral arms on the large scale (outside Hill radius of the binary indicated by outer white circles) are smoothed out as they are not stationary in the binary's rotating frame.

We find that the torque density can be well described by four distinctive quadrants. It is positive in the first and third quadrants, while negative in the second and fourth quadrants. By comparing with the stacked surface density plots, both the inter-spiral arms and the outer-spiral arms contribute negative to the total gravitational torque, while the two-tips of the outer-spiral arms located in the second and fourth quadrants contribute positive torque. 

When the softening is small, the positive contribution from the CSD region dominates over the negative contribution from the spiral arms, as shown from the symmetric torque density plot (the upper right panel of Figure~\ref{fig:torqden}), and the radial cumulative torque shown in Figure~\ref{fig:torq_r}.
Inside of $0.5\ a_0$, the interconnecting spiral torques the binary down, while between $0.5\ a_{0}$ and $a_{0}$ the CSD spiral arms add enough angular momentum to the binary to overcome the negative torque from inside $0.5\  a_{0}$. 
Outside $a_{0}$, the CBD spirals do not contribute enough negative torque to prevent the binary from expanding.

However, as the softening grows, the positive contribution of the CSD weakens while the negative contribution of the CBD grows. 
From the density plot, we see that larger softening washes out the CSD spirals.
Since these spirals are the predominant source of positive torque, larger softening results in contracting binaries due to the relatively large CBD torque.

\subsection{Binary Eccentricity and Destruction}\label{sec:instability}

An intriguing feature for the outspiraling binary is the rapid increase in eccentricity after its semi-major axis exceeds $\sim 0.04\ r_{0}$. 
The maximum eccentricity during this stage can be as large as 0.5, as shown in the second panel of Figure~\ref{fig:rbin}.  Since the semi-major axis of the binary $a_{\rm bin}$ does not increase significantly,  the minimum separation can be still be smaller than the initial value of $a_{0}$. 

If we calculate $\dot{e}_{\rm bin}$ due to the disk, the rate we find is about one order of magnitude smaller than what is shown in Figure~\ref{fig:rbin}.
This suggests that the rapid increase in binary eccentricity is {\it not} due to the disk, but to the interaction with the central SMBH.
Indeed, there are BBH-destroying instabilities in hierarchical co-planar triple systems that occur at our chosen masses when $a_{\rm bin} \sim 0.04\ a_{\rm com} \approx 0.36\ R_{\rm H}$ \citep{Eggleton1995,Mardling2001}.
This threshold for instability roughly agrees with the semi-major axis at which our binaries experience rapid eccentricity growth and eventual destruction. The slightly larger threshold compared to the theoretical expectation could be due to the existence of disk, which leads to the onset of instability more gently.

\subsection{BBH Accretion} \label{sec:accretion}

\subsubsection{Steady-state accretion}

In steady-state the time and azimuthally averaged accretion rate through the disk should be constant. 
For an accreting BBH, the accretion rates in the disk interior ($r<a_{\rm com}$) and exterior ($r>a_{\rm com}$) to the BBH COM will be different but satisfy $\dot{m}_{\rm in} - \dot{m}_{\rm outer} = \dot{m}_{\rm bin}$.
Here a positive $\dot{m}_{\rm bin}$ denotes mass removal and a positive $\dot{m}_{\rm in},\dot{m}_{\rm outer}$ refers to outwards mass transport.
We find that $\dot{m}_{\rm bin}$ depends on the adopted removal rate and that if this rate is too fast the global flow of the disk can be reversed, i.e., $\dot{m}_{\rm in}$ can be outwards rather than inwards.

\begin{figure}
\centering
\includegraphics[width=0.48\textwidth,clip=true]{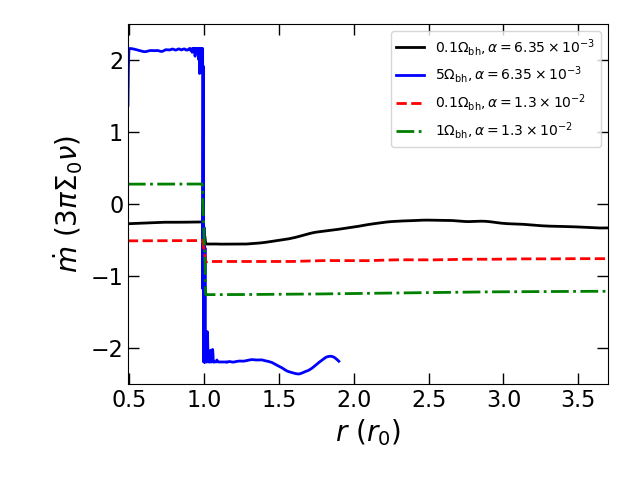}
\caption{Disk accretion profiles for different removal rates and disk viscosities. The blue line shows the azimuthal and time averaged (between $400-500$ orbits) disk accretion rate before the binary release for Model A in Table~\ref{tab:para} as a function of radial distance from the central SMBH (Note that the wave damping boundary layers have been cut off.). The black line corresponds to Model Ar in Table~\ref{tab:para}. The jump at $r=r_{0}$  matches exactly the accretion rate onto the BBH shown in Figure~\ref{fig:mdot}. The other two lines are with a higher disk viscosity ($\alpha=1.3\times10^{-2}$) and two removal rates as indicated by the legends. We can see that a large removal rate results in the inner disk accretion rate becomes unphysically positive.  
}
 \label{fig:mdisk}
\end{figure}

We show this effect in Figure \ref{fig:mdisk} by plotting the time-averaged $\dot{m}(r)=  \int d\phi r  \Sigma v_r$ for two different viscosities and domain sizes. 
The time-averaging is done over 100 COM orbits at a time just before the BBH is released.
For removal rates of $\tau_{\rm acc}^{-1} = \Omega_{{\rm bh},i}$ and $5 \Omega_{{\rm bh},i}$, $\dot{m}_{\rm in} > 0$. 
In these situations, $\dot{m}_{\rm bin} > |\dot{m}_{\rm outer}|$ , i.e the BBH wants to accrete faster than the outer disk can feed it and so the inner disk must compensate. 
For a slower removal rate of $0.1 \Omega_{{\rm bh},i}$, the disk maintains inwards mass transport everywhere and maintains $\dot{m}_{\rm bin} < |\dot{m}_{\rm outer}|$.  
In particular, we find that just before release the BBH accretes $\sim 50\%$ of $\dot{m}_{\rm outer}$ for slow removal rates.

\begin{figure}[htbp]
\centering
\includegraphics[width=0.45\textwidth,clip=true]{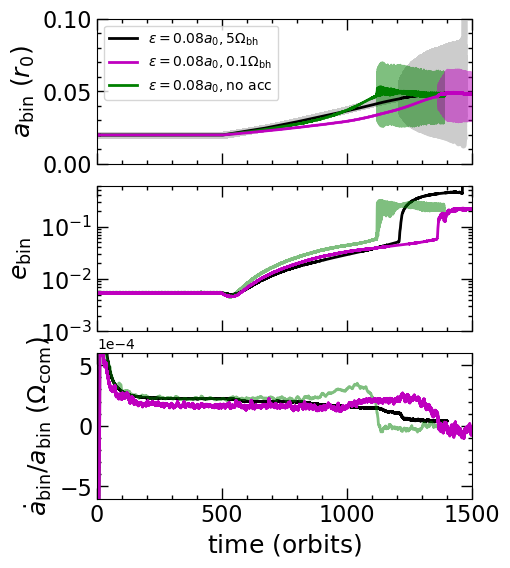}
\caption{
Same as Figure \ref{fig:rbin}, but for different accretion methods.
Shown are Models A (black),  Ar (magenta), and D (green). 
All three models show similar binary dynamics.
}
 \label{fig:rbin_tacc}
\end{figure}

\subsubsection{Time evolution}

While the global accretion rate in the disk sensitively depends on the removal rate, the time evolution of the BBH itself does not. 
This is shown in Figure~\ref{fig:rbin_tacc} which plots the evolution of the BBH for different accretion prescriptions.
Despite the vastly different $\dot{m}(r)$ and $\dot{m}_{\rm bin}$, at early times the BBH expands and gains eccentricity at roughly the same rate -- indicating that these large scale effects are unimportant near the BBH.
In fact, we find that the flow and spatial torques on the CSD scale are roughly the same. 
If we were to take into account, the additional mass and angular momentum accreted by the binary, this result may change, but simulations of isolated binaries find this component of the torque is typically small \citep{Munoz2019,Moody2019}.

\begin{figure}
\centering
\includegraphics[width=0.48\textwidth,clip=true]{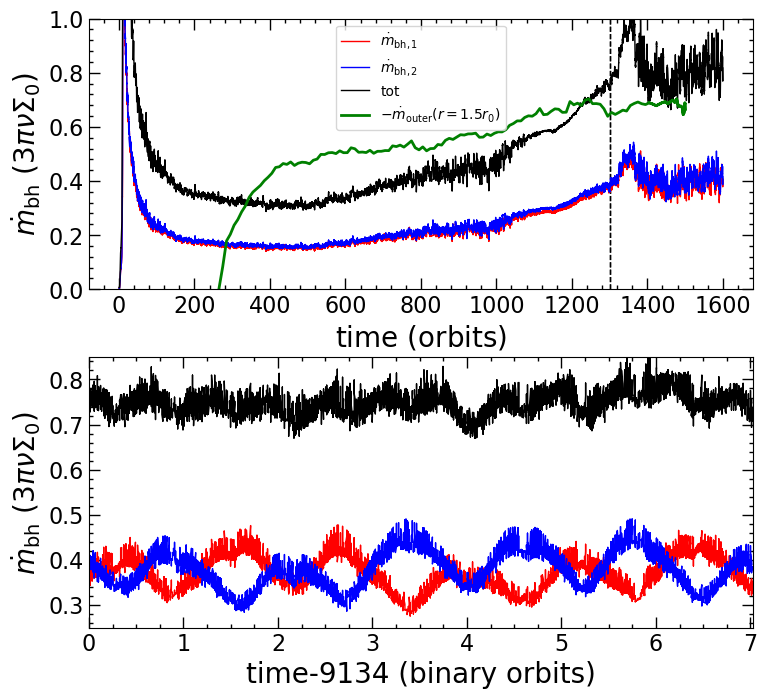}
\caption{BBH accretion rates for Model Ar.  Upper panel: time-averaged BBH accretion rates. Red and blue lines correspond to accretion rates for two members of the binary,  and the black line is their sum. The green line shows the time evolution of the supplied accretion rate from the outer disk (i.e., $-\dot{m}_{\rm outer}$ at $r=1.5\ r_{0}$). 
The lower panel shows a zoom-in of the accretion rates without averaging between 1300 and 1301 orbits when the binary eccentricity is $\approx 0.06$.
}
 \label{fig:mdot}
\end{figure}

Figure \ref{fig:mdot} shows the time evolution of $\dot{m}_{\rm bin}$ and the accretion rates of each individual BH for the slow removal case.
At early times, the binary accretes symmetrically with $\dot{m}_{\rm bin} \sim [0.5-0.8] |\dot{m}_{\rm outer}|$, with a periodic variation of $2\Omega_{\rm bin}$, where $\Omega_{\rm bin}$ is the current binary orbital frequency.

At later times when the binary has become eccentric, we find that one component of the binary preferentially accretes more material than the other on a $\sim 2$ binary orbit timescale.
This periodic "flipping" is also seen is simulations of isolated eccentric binaries \citep[e.g.,][]{Dunhill2015,Munoz2016,Munoz2019}.
However, unlike the isolated binary simulations we do not find any accretion variability on a $5$ orbit timescale \citep[see e.g.,][]{MacFadyen2008,Shi2012,DOrazio2013,Farris2014,Munoz2016,Miranda2017,Munoz2019,Duffell2020}. 
One common explanation for the $\Omega_{\rm bin}/5$ variability is the build up of a gas lump at the edge of the binary cavity that subsequently drains onto the binary every five orbits.
We suspect that the absence of this variability is due either to the BBH CBD being fed from the large scale spirals, as opposed to an axisymmetric viscously controlled inflow, or the lack of a clear cavity surrounding the BBH.

Only at later times -- well after the binary is released -- do we see an appreciable difference between slow and fast accretion. 
In particular, at $t=1250$ COM orbits, there is a noticeable slowdown in the BBH expansion rate for our slow accretion model. 
We suspect that this could be related to the rapid increase of binary eccentricity,
as simulations of isolated binaries have found that the expansion rates for eccentric binaries are slower than their circular counterparts \citep{Munoz2019}.

\subsection{Other Models: Influence of Binary Initial Separation and Gap Depth} \label{sec:other_models}

\begin{figure}[t]
\centering
\includegraphics[width=0.45\textwidth,clip=true]{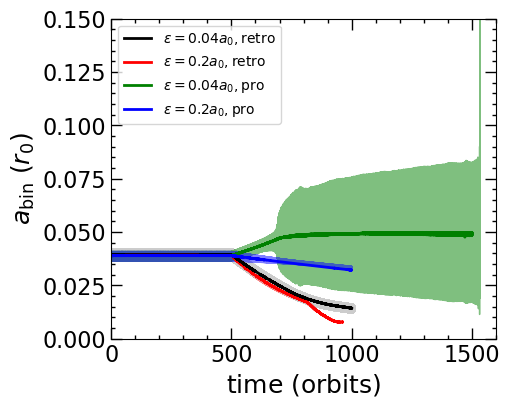}
\caption{Binary separation (shaded regions) and semi-major axis (solid lines) evolution for models with an initial separation of $a_{0}=0.04\ r_{0}$. Green ($\epsilon=0.04\ a_{0}$, Model E) and blue ($\epsilon=0.2\ a_{0}$, Model F) lines are for prograde orbits and black ($\epsilon=0.04\ a_{0}$, Model G) and red ($\epsilon=0.2\ a_{0}$, Model H) lines are retrograde  orbits. Note that for the retrograde orbits both small and large softening cases cause the binary orbits contract with time.
}
 \label{fig:rbin_dr_ret}
\end{figure}

\paragraph{Initial separation}
We have another two runs with the initial separation of the binary of $a_{0}=0.04\ r_{0}$ (Models E, F). The binary evolution profiles are shown in Figure~\ref{fig:rbin_dr_ret}. Similar to the Models A \& B shown in Figure~\ref{fig:rbin}, two softenings with $\epsilon=0.04\ a_{0}$ and $\epsilon=0.2\ a_{0}$ have been tested. We obtain very similar results as the $a_{0}=0.02\ r_{0}$ cases: the binary is outspiraling when $\epsilon=0.04\ a_{0}$ and it is inspiraling when $\epsilon=0.2\ a_{0}$, as long as the initial separation is less than the critical value required to trigger the instability for the triple system discussed in Section~\ref{sec:instability}.

\paragraph{Gap depth and $H/R_{\rm H}$}
Our Model A in Table~\ref{tab:para} has a gap depth of $\Sigma_{\rm BBH}\simeq0.03\Sigma_{0}$ and $R_{\rm H}\gtrsim H$. We have tested several other cases with smaller gap depths 
and different binary embeddedness  parameters $R_{\rm H}/H$, by changing $\alpha$ and $h_{0}$. 
Table \ref{tab:para} shows $\dot{a}_{\rm bin}/a_{\rm bin}$ for each of these setups.
We find that BBHs expand irrespective of gap depth or $R_{\rm H}/H$, but we note that as the binary becomes more embedded the expansion rate slows down.
This is different than what was found by \citet{Tiede2020} for isolated binaries, as those authors found that binaries eventually harden as $h_{0}$ decreases.
Our 2D simulations, however, may not be appropriate for the deeply embedded case (see Section \ref{sec:future} for further discussion).

\begin{table*}[t]
  \begin{center}
  \caption{\bf Binary evolution $\dot{a}_{\rm bin}/a_{\rm bin}$ for different disk parameters}\label{tab:para_disk0}
  \begin{tabular}{l|ccc}
     \hline\hline
     % after \\: \hline or \cline{col1-col2} \cline{col3-col4} ...
      & $h_{0}=0.05$ & $h_{0}=0.08$ &  $h_{0}=0.15$  \\
      & $(R_{\rm H}>H)$ & $(R_{\rm H}\gtrsim H)$ & $(R_{\rm H}< H)$  \\
      
     \hline
     
     $\Sigma_{\rm BBH}\gtrsim 0.5\Sigma_{0}$ & -- & $1.1\times10^{-3}$ & -- \\
      No gap & -- & ($\alpha=0.2$) & -- \\
      
      \hline
     $0.1 \Sigma_0\lesssim \Sigma_{\rm BBH}\lesssim 0.5 \Sigma_0$ & $1.0\times10^{-3}$ & $5.0\times10^{-4}$ & $7.5\times10^{-5}$ \\
      Partial gap & ($\alpha=0.33$) & ($\alpha=0.032$) &  ($\alpha=1.3\times10^{-3}$) \\
      
      \hline

      $\Sigma_{\rm BBH}\lesssim0.1\Sigma_{0}$ & $5.0\times10^{-4}$ & $2.5\times10^{-4}$ & $5.0\times10^{-5}$ \\
      Deep gap & ($\alpha=0.066$) & (Model A) & ($\alpha=2.7\times10^{-4}$) \\     
     
    \hline\hline
   \end{tabular}
   \end{center}
   
   \tablecomments{All other disk parameters are the same as Model A in Table~\ref{tab:para} except for $\alpha$ and $h_{0}$ stated in this Table. All values for $\dot{a}_{\rm bin}/a_{\rm bin}$ are in unit of $\Omega_{\rm com}$. A positive $\dot{a}_{\rm bin}/a_{\rm bin}$ means that the BBH expands with time. \\
   }
  
\end{table*}

\begin{figure}[t]
\centering
\includegraphics[width=0.45\textwidth,clip=true]{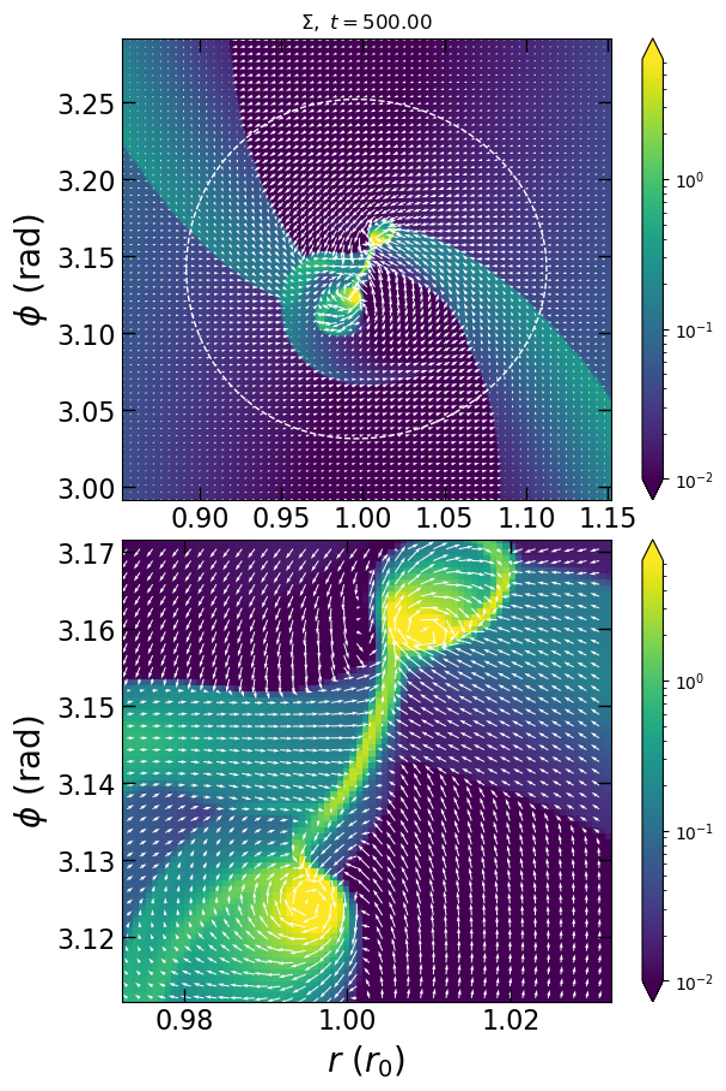}
\caption{Same as Figure~\ref{fig:sigma_snapshot} but for Model G. The global gas surface density is similar to that in Figure~\ref{fig:sigma_snapshot}.
}
 \label{fig:sigma_snapshot_ret}
\end{figure}

\subsection{Retrograde Binary}

Up until now, we have only focused on prograde binaries. 
Here we further examine the impact of retrograde orbits, i.e, when the BH-BH angular momentum is mutually inclined to that of the BBH-COM by $180^\circ$. 
These are Models G and H in Table~\ref{tab:para}.

We show a close-up of the gas surface density and velocity for Model G at 500 orbits in Figure~\ref{fig:sigma_snapshot_ret}. 
The global gas surface density is quite similar to the prograde case shown in Figure~\ref{fig:sigma_snapshot}. 
However, the gas distribution and flow pattern in the vicinity of the BBH are quite complex.
There are clear CSDs that are aligned with the BBH's angular momentum vector, but the spiral pattern is much less pronounced than in the prograde case.

We show the semi-major axis evolution for Models G and H in Figure \ref{fig:rbin_dr_ret}.
Interestingly, the binary is inspiraling even for a small softening.
The inspiraling rate for the retrograde orbit is larger than that of the large softening, prograde case \citep{Baruteau2011}. From the separation evolution, the binary remains circular as it shrinks.

\begin{figure}[t]
\centering
\includegraphics[width=0.45\textwidth,clip=true]{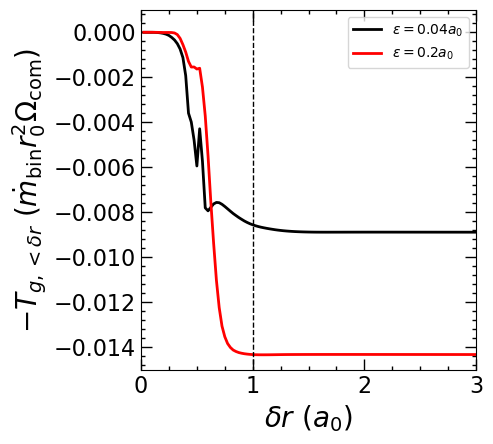}
\caption{Same as Figure\ \ref{fig:torq_r} but for the retrograde orbits with an initial separation of $0.04\ r_{0}$ for the binary. Dashed line indicates the CSD region with $\delta r \le a_{0}$. Note that $-dT_{{\rm g},<\delta r}$ is shown here for the retrograde cases.}
\label{fig:torq_r_ret}
\end{figure}

\begin{figure*}[t]
\centering
\includegraphics[width=0.98\textwidth,clip=true]{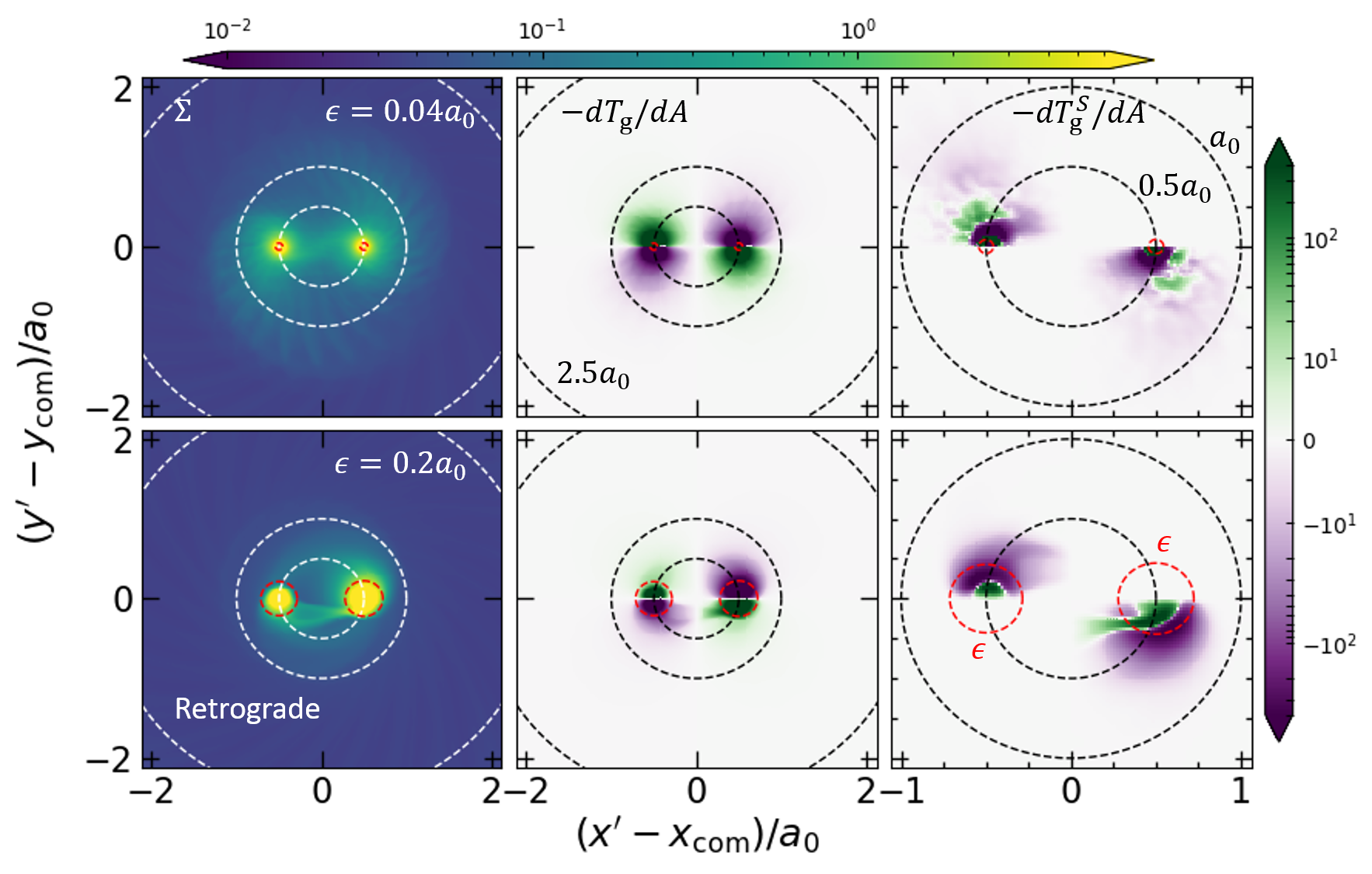}
\caption{Same as Figure\ \ref{fig:torqden} but for Models G \& H which have retrograde orbits. The upper panels are for the softening of $\epsilon=0.04\ a_{0}$ (Model G), the lower panels are for $\epsilon=0.2\ a_{0}$ (Model H).  Note that a minus sign is added to the torque density in the middle and right panels for the retrograde cases.
}
 \label{fig:torqden_ret}
\end{figure*}

As we did for the prograde case, we show the cumulative torque and stacked, 2D torque density in Figures \ref{fig:torq_r_ret} and \ref{fig:torqden_ret}.
The surface density and torque density are averaged over 400 snapshots between $500-504\ P_{\rm com}$.
Note that because the angular momentum vector of the binary is flipped, we flip the sign of the torque so that positive values still correspond to increasing $a_{\rm bin}$.

From Figure \ref{fig:torq_r_ret}, we see that retrograde BBHs inspiral regardless of softening. 
In contrast to the prograde case, the CSD region actually makes the binary inspiral at a faster rate.
By looking at Figure \ref{fig:torqden_ret}, we see that this is because the CSDs lack any coherent spiral structure that can provide an outspiraling torque.

\section{Conclusions and Discussion} \label{sec:conclusions}

We have performed a series of global 2D isothermal hydrodynamical simulations of an equal mass-ratio BBH embedded in an accretion disk to study the evolution of the binary separation.

We have shown that prograde binaries, which are most often discussed in the literature, {\it expand} rather than contract, as is often assumed.
This is because when the CSD region is adequately resolved -- with both high resolution and small softening -- the positive CSD contribution to the total torque overwhelms the negative contribution from the CBD and inter-spiral regions.
Therefore, a small softening with a high resolution to properly resolve the CSD region is crucial, in agreement with recent isolated binary simulations \citep{Moody2019,Munoz2019,Munoz2020,Duffell2020,Tiede2020}.
Such a conclusion does not sensitively depend on the accretion prescription for the binary we adopt.
These results caution against the popularity of the AGN disk channel for prograde BBH mergers as suggested in many previous works \citep{McKernan2012,McKernan2014,McKernan2018,McKernan2019,Bartos2017,Stone2017,Leigh2018,Secunda2019,Secunda2020a,Yang2019a,Yang2019b,Grobner2020,Ishibashi2020,LigoVirgo2020,Tagawa2020,Tagawa2020b}.

However, when the softening scale is about half of the initial binary separation (i.e., $\epsilon=0.4\ a_{0}$), the binary orbit contracts with time and stalls at the softening scale, consistent with \citet{Baruteau2011}. 
The stalling of the orbital hardening at this large softening scale $\epsilon$ thus cannot bring the binary into a gravitational-wave (GW)-emitting dominated regime.

In contrast to the prograde case, we find that retrograde binaries do contract, regardless of softening.
This is due primarily to the loss of the spiral structure in the CSDs that provides the main source of expansion.
This suggests that the gas-driven inspiral for retrograde binaries may produce a population of compact BBHs in the GW-emitting regime in AGN disks, which may contribute a large fraction to the observed LIGO/Virgo events
\citep[as suggested by e.g.,][]{Secunda2020a}.

For the outspiraling case, the binary's eccentricity can be significantly excited within a few hundred binary COM orbits. Due to the eccentricity growth, the minimum separation between the binary could still be smaller than its initial value, both of which can enhance the GW inspiraling rate significantly due to the strong dependence of the coalescence time scale on the binary's eccentricity (i.e., $t_{\rm gw}\propto(1-e_{\rm bin})^{3.5}$). We expect that the the maximum binary's eccentricity could be even higher if we decrease the initial semi-major axis of the binary. If this is the case, a highly eccentric BBH merger could be observed by future GW observations \citep{Gayathri2020,Samsing2020}. It is also possible that the binary will be destroyed due to the dynamical instability of the triple system, if they have not entered into the GW-emitting regime yet after the rapid eccentricity growth stage.

\subsection{Limitations and Future Prospect} \label{sec:future}

In this work we assume a small gravitational softening by analogy with the isolated binary simulations.
However, simulations of single intermediate mass ratio secondaries use a much larger softening of $\epsilon\simeq0.7\ H$.
This number has been shown to produce approximately equal 2D and 3D gravitational torques \citep{Muller2012}.
However, for a 2D BBH simulation, this softening is more appropriate for scales larger than $a_{\rm bin}$, i.e., to obtain an equivalent 3D torque on the binary's COM outside of $\delta r \sim H$, but certainly not for smaller scales where the dynamics are very different from that of a single object.
A study similar to \citet{Muller2012} which compares the 2D and 3D results for embedded BBHs is thus required to determine the ``correct" softening scale. 
We suspect that the answer will actually be two softening lengths, one for the larger scale that is related to the COM potential, and one for the smaller scale that is related to the higher moments of the BBH potential.
Further work, however, is required.

The dependence of our result on the inclination of the BBH to the central AGN remains unknown. 
We have shown with 2D simulations that there is a transition from expansion at $0^\circ$ inclination to contraction $180^\circ$ inclination.
We would then expect that there is a critical inclination where BBHs begin to harden.
This transition may likely occur at inclinations greater than $45^\circ$, as recent work has shown that isolated binaries inclined by $45^\circ$ still expand \citep{Moody2019}.
Once inclined, the BBH will also excite warps \citep{Fragner2010,Foucart2014,Moody2019} in the disk and may even break the disk -- especially at the mass ratios we consider \citep[as shown in e.g.,][]{Zhu2019}. 
It is unclear how these processes affect the BBH's evolution and will require dedicated 3D simulations.

Our main set of simulations consider BBHs with $R_{\rm H}\gtrsim H$.
For these systems, we would expect true CSDs to form as most of the accretion flow is in the plane of the binary. 
However, 3D simulations of planet-disk interaction at comparable masses to ours show that some material is accreted from higher latitudes \citep{Fung2016,Szulagyi2014}.
Capturing these 3D flows may be important for determining the BBH accretion rate.

Depending on the AGN disk temperature and BH masses, BBHs may actually be embedded in the disk, i.e. have $R_{\rm H} < H$. 
In this case, our 2D treatment of the dynamics becomes a poor approximation.
In particular, if fully embedded BBHs cannot form CSDs and instead accrete from an envelope \citep{Szulagyi2016,Szulagyi2017}, the important repulsive torques in this region may be absent, and the BBHs may harden.

In all of our simulations, we have adopted a locally isothermal equation of state, which could be unrealistic, especially within the region of each BH's Hill  sphere. 
We intend to implement a more detailed treatment of the thermodynamics and/or more a realistic viscosity/temperature prescription around each BH to take account into the radiative cooling/heating of the gas. 
Higher temperatures in the Hill region may reduce the strength of the CSD spirals, and consequently their torques, or prevent the formation of CSDs entirely \citep{Szulagyi2016,Szulagyi2017}.
Furthermore, a more accurate treatment of accretion onto each BH in this region, and any feedback onto the disk, may alter the structure of the CSDs \citep[as shown in e.g.,][]{SouzaLima2017,delValle2018}.
However, it is unclear how well the feedback results of these studies, which focus on isolated supermassive BBHs, as well as the results from single AGN feedback studies \citep[e.g.,][]{Yuan2018,Yoon2018,Li2018} extrapolate to embedded stellar mass BBHs.

Finally, there are various parameter spaces to be explored in the future.  This includes the binary mass ratio, the mass ratio to the central SMBH, the disk aspect ratio, and disk mass.
We should expect the binary to inpiral for low mass ratio based on the consensus of inward migration for a giant planet in protoplanetary disks. The transition from outspiraling to inspiraling for a decreasing mass ratio in isolated binary simulation has been reported recently (\citealt{Duffell2020}, Dempsey et al. 2020, in prep.). This will lead to an extreme mass ratio inspiral event from the BBH merger.
We have explored  a regime with the disk mass being less than the mass of BBH, but the systems with more massive disk could behave very differently  \citep{Cuadra2009,Roedig2012}, which may also drive the binary to coalescence.

\acknowledgments
We thank the referee for useful comments. We would like to thank Douglas Lin, Cl\'{e}ment Baruteau, Bin Liu, and Barry McKernan for beneficial discussions. A.M.D thanks Diego Mu\~{n}oz for many helpful discussions on binary-disk interaction. 
We gratefully acknowledge the support by LANL/LDRD and NASA/ATP. This research used resources provided by the Los Alamos National Laboratory Institutional Computing Program, which is supported by the U.S. Department of Energy National Nuclear Security Administration under Contract No. 89233218CNA000001. 
The LANL Report number is LA-UR-20-28080.
Softwares: \texttt{LA-COMPASS} \citep{Li2005,Li2009}, \texttt{Numpy} \citep{vanderWalt2011}, \texttt{Matplotlib} \citep{Hunter2007}

\bibliography{references.bib}{}
\bibliographystyle{aasjournalnolink}

\end{document}